\documentclass[aps,pra,twocolumn,floatfix,bibnotes,superscriptaddress]{revtex4-2}
\usepackage{graphicx}
\usepackage{amsmath}
\usepackage{amssymb}
\usepackage{bm}
\usepackage{color}
\usepackage[scr=boondoxo]{mathalfa}
\usepackage{hyperref}
\usepackage{float}

\begin{document}

\title{Blackbody radiation Zeeman shift in Rydberg atoms}

\newcommand{\NIST}{
National Institute of Standards and Technology, Boulder, Colorado, USA
}

\newcommand{\CU}{
Department of Physics, University of Colorado, Boulder, Colorado, USA
}

\author{K. Beloy}
\email{kyle.beloy@nist.gov}
\affiliation{\NIST}

\author{B. D. Hunt}
\affiliation{\NIST}
\affiliation{\CU}

\author{R. C. Brown}
\affiliation{\NIST}

\author{T. Bothwell}
\affiliation{\NIST}

\author{Y. S. Hassan}
\affiliation{\NIST}
\affiliation{\CU}

\author{J. L. Siegel}
\affiliation{\NIST}
\affiliation{\CU}

\author{T. Grogan}
\affiliation{\NIST}
\affiliation{\CU}

\author{A. D. Ludlow}
\affiliation{\NIST}
\affiliation{\CU}

\date{\today}

\newcommand{\later}{\ensuremath{\spadesuit}}
\newcommand{\pprime}{\ensuremath{{\prime\prime}}}
\newcommand{\e}{\ensuremath{\mathscr{e}}}
\newcommand{\codata}{\mbox{CODATA}} % prevents CODATA from being split across lines

\begin{abstract}
We consider the Zeeman shift in Rydberg atoms induced by room-temperature blackbody radiation (BBR). BBR shifts to the Rydberg levels are dominated by the familiar BBR Stark shift. However, the BBR Stark shift and the BBR Zeeman shift exhibit different behaviors with respect to the principal quantum number of the Rydberg electron. Namely, the BBR Stark shift asymptotically approaches a constant value given by a universal expression, whereas the BBR Zeeman shift grows steeply with principal quantum number due to the diamagnetic contribution. We show that for transitions between Rydberg states, where only the differential shift between levels is of concern, the BBR Zeeman shift can surpass the BBR Stark shift. We exemplify this in the context of a proposed experiment targeting a precise determination of the Rydberg constant.
\end{abstract}

\maketitle

\section{Introduction}

Rydberg atoms have special features that make them appealing for a diverse range of applications. These features include long lifetimes, strong interatomic interactions, enhanced sensitivity to external fields, and theoretical tractability, while the applications include quantum information processing, quantum simulation, electromagnetic field sensing, and fundamental physics~\cite{Gal94,SafWalMol10,JonMarSha17,AdaPriSha19,BroThi20}. For instance, Ramos~{\it et al.}~\cite{RamMooRai17} have proposed a Rydberg-atom experiment that aims to determine the Rydberg constant effectively independent of additional experimental input, with uncertainty approaching that of the \codata{} recommended value. The \codata{} evaluation of the Rydberg constant incorporates a multitude of experimental data, with the Rydberg constant being strongly coupled to the proton charge radius. In 2010, the proton charge radius was independently measured using muonic hydrogen, resulting in substantial tension with the then-recommended value~\cite{PohAntNez10}. Not only was the true value of the proton charge radius brought into question, but so to was the true value of the Rydberg constant and even the legitimacy of the standard model~\cite{PohGilMil13}. With additional experimental data, the ``proton radius puzzle'' is now deemed at least partially resolved, with the latest \codata{} recommended values for the proton charge radius and Rydberg constant being substantially shifted relative to their prior recommended values~\cite{TieMohNew21,CODATA2022online}. In any case, it is clear that obtaining an independent measure of the Rydberg constant is a worthy endeavor, and precision spectroscopy with Rydberg atoms offers a viable path towards this goal.

Blackbody radiation (BBR), with its inherent electric field, induces Stark shifts to atomic energy levels~\cite{GalCoo79,FarWin81,ItaLewWin82}. The ground state of Rb, for instance, experiences a shift of $-2.8$~Hz~\cite{GreHroHol15} due to room-temperature BBR. This is the typical size of the shift for low-lying states of open-shell, neutral atomic systems.
%For the lowest-lying states in neutral atoms, room-temperature BBR Stark shifts are typically on the order of Hz. For state-of-the-art atomic clocks, for example, the corresponding shift to the clock transition frequency is notoriously one of the largest and most challenging systematic frequency shifts, necessitating specialized apparatus and requiring careful characterization~\cite{LudBoyYe15,UshTakDas15,BelHinPhi14,HeoKimYu22,AepKimWar24}.
For Rydberg states, on the other hand, the level shifts are expected to be enhanced by a few orders of magnitude. Notably, theory predicts that the BBR Stark shift for high-lying Rydberg states is given, at least approximately, by a simple universal expression, evaluating to $\approx\!2.4$~kHz at room temperature~\cite{FarWin81}. An immediate implication is that, while the shifts to the levels may be large, shifts to transition frequencies may be relatively suppressed for transitions between Rydberg states.

In addition to its inherent electric field, BBR also has an inherent magnetic field. Consequently, BBR also induces Zeeman shifts to atomic energy levels. For low-lying states in neutral atoms, room-temperature BBR Zeeman shifts are small ($\lesssim\!10~\mu$Hz), being negligible even for state-of-the-art atomic clocks~\cite{PorDer06}. Diamagnetism, however, is known to be amplified in Rydberg states, usually being considered in the context of dc magnetic fields~\cite{Gal94}. Diamagnetism originates from a perturbative term in the Hamiltonian proportional to the amplitude-squared of the electromagnetic vector potential (see Sec.~\ref{Sec:derive} below). It induces level shifts quadratic in the magnetic field amplitude and, as such, can be anticipated to contribute to BBR shifts.

Starting from first principles, here we consider the BBR Zeeman shift in Rydberg atoms. In particular, we demonstrate that for transitions between Rydberg states, where only the differential shift between levels is pertinent, the BBR Zeeman shift can surpass the BBR Stark shift. This result is attributed to the enhanced diamagnetic contribution to the BBR Zeeman shift, as well as the appreciable cancellation in the BBR Stark shift. We conclude that the BBR Zeeman shift may need to be taken account in future precision-spectroscopy experiments based on Rydberg atoms.

Room-temperature BBR is assumed throughout unless otherwise stated. When not explicitly specified, a temperature of 300~K is used for numerical evaluations, while temperature dependence can be appreciated from the unevaluated expressions. SI expressions are employed throughout. Physical constants appearing in expressions below include the fine structure constant ($\alpha$), Planck's constant ($h$), Planck's reduced constant ($\hbar$), the speed of light ($c$), the elementary charge ($e$), the electron mass ($m$), the vacuum permittivity ($\varepsilon_0$), the Bohr radius ($a_B$), and the Bohr magneton ($\mu_B$). Mathematical constants appearing in the expressions below include the imaginary unit ($i$) and the base of the natural logarithm ($\e$, not to be confused with $e$).

\section{BBR Stark and BBR Zeeman shifts in Rydberg atoms}
\label{Sec:derive}

We employ a single-active-electron model for the Rydberg atom, with nuclear spin being neglected. The motion of the (Rydberg) electron is treated nonrelativistically. For simplicity, we assume a neutral alkali-metal atom, though conclusions are expected to hold more broadly (e.g., in absence of effects from perturber states~\cite{Fan75,Sea83,AymGreLuc96}). We begin with the Hamiltonian
\begin{align}
H={}&
\frac{\left[\mathbf{p}+e\mathbf{A}\left(\mathbf{r},t\right)\right]^2}{2m}
-e\phi\left(r\right)
-e\Phi\left(\mathbf{r},t\right)
\nonumber\\
&-\bm{\mu}_s\cdot\left[\bm{\nabla}\times\mathbf{A}\left(\mathbf{r},t\right)\right],
\label{Eq:startHam}
\end{align}
where $\mathbf{r}$, $\mathbf{p}$, and $\bm{\mu}_s$ are the position relative to the nucleus, the conjugate momentum, and the spin magnetic moment, respectively, of the electron. The spin magnetic moment can further be written as $\bm{\mu}_s=-\left(g_s\mu_B/\hbar\right)\mathbf{s}$, where $g_s\approx2$ is the electron $g$-factor and $\mathbf{s}$ is the spin angular momentum of the electron. $\phi\left(r\right)$ is a time-independent, spherically symmetric scalar electromagnetic potential attributed to the atomic core (nucleus + core electrons); it is understood to approach $\phi(r)\rightarrow\left(e/4\pi\varepsilon_0\right)(1/r)$ outside of the core region. $\Phi(\mathbf{r},t)$ and $\mathbf{A}(\mathbf{r},t)$ are scalar and vector electromagnetic potentials associated with external fields, which generally depend on space and time. Finally, $\left[\bm{\nabla}\times\mathbf{A}\left(\mathbf{r},t\right)\right]$ in Eq.~(\ref{Eq:startHam}) is the curl of the vector potential evaluated at the position of the electron, with $\bm{\nabla}$ having no effect outside of the brackets.

We first consider the case of static, uniform external fields. In this case, the electromagnetic potentials may be taken as
\begin{gather*}
\Phi\left(\mathbf{r},t\right)=-\mathbf{E}\cdot\mathbf{r},
\\
\mathbf{A}\left(\mathbf{r},t\right)=\frac{1}{2}\left(\mathbf{B}\times\mathbf{r}\right),
\end{gather*}
where $\mathbf{E}$ and $\mathbf{B}$ are the electric and magnetic fields, respectively. Inserting these expressions into equation~(\ref{Eq:startHam}) and performing a few intermediate steps, we can express the Hamiltonian as $H=H_0+V$, with
\begin{gather}
H_0=\frac{p^2}{2m}-e\phi(r),
\nonumber\\
V=-\mathbf{E}\cdot\mathbf{d}-\mathbf{B}\cdot\bm{\mu}+\frac{e^2\left|\mathbf{B}\times\mathbf{r}\right|^2}{8m}.
\label{Eq:Vstatuni}
\end{gather}
Here $\mathbf{d}=-e\mathbf{r}$ and $\bm{\mu}=\bm{\mu}_\ell+\bm{\mu}_s$ are the electric dipole and magnetic dipole operators for the electron. The magnetic dipole includes an orbital contribution $\bm{\mu}_\ell=-(\mu_B/\hbar)\bm{\ell}$, where $\bm{\ell}=\mathbf{r}\times\mathbf{p}$ is the orbital angular momentum of the electron, in addition to the spin contribution discussed above. The last term in $V$ is the diamagnetic interaction.

For more general fields $\mathbf{E}\left(\mathbf{r},t\right)$ and $\mathbf{B}\left(\mathbf{r},t\right)$ depending on both space and time, the Hamiltonian may still be expressed as $H=H_0+V$, with $H_0$ as given above. Meanwhile, $V$ can be approximated by Eq.~(\ref{Eq:Vstatuni}) with the static, uniform fields being replaced by their generalized counterparts evaluated at the atomic center, $\mathbf{E}\rightarrow\mathbf{E}\left(0,t\right)$ and $\mathbf{B}\rightarrow\mathbf{B}\left(0,t\right)$. We will be content with this approximation, which omits higher multipolar contributions and retardation effects~(see, e.g., Ref.~\cite{AnzShiSat18}).

For the case of a plane electromagnetic wave, the fields can be written as
\begin{gather*}
\mathbf{E}\left(\mathbf{r},t\right)=\frac{\mathcal{E}}{2}
\left[\bm{\epsilon}\e^{i\left(\mathbf{k}\cdot\mathbf{r}-\omega t\right)}
+\bm{\epsilon}^*\e^{-i\left(\mathbf{k}\cdot\mathbf{r}-\omega t\right)}
\right],
\\
\mathbf{B}\left(\mathbf{r},t\right)=\frac{\mathcal{B}}{2}
\left[\bm{\beta}\e^{i\left(\mathbf{k}\cdot\mathbf{r}-\omega t\right)}
+\bm{\beta}^*\e^{-i\left(\mathbf{k}\cdot\mathbf{r}-\omega t\right)}
\right],
\end{gather*}
where $\omega$ is the angular frequency and $\mathbf{k}=(\omega/c)\bm{\kappa}$ is the wave vector. $\mathcal{E}$ and $\bm{\epsilon}$ are the amplitude and polarization vector associated with the electric field, while $\mathcal{B}=\mathcal{E}/c$ and $\bm{\beta}=\bm{\kappa}\times\bm{\epsilon}$ are the magnetic analogs. $\bm{\kappa}$, $\bm{\epsilon}$, and $\bm{\beta}$ are unit vectors satisfying $\bm{\kappa}\cdot\bm{\epsilon}=\bm{\kappa}\cdot\bm{\beta}=\bm{\epsilon}\cdot\bm{\beta}=0$, with real $\bm{\kappa}$ and generally complex $\bm{\epsilon}$ and $\bm{\beta}$. The asterisk indicates complex conjugation. Writing $V=V_{E1}+V_{M1}+V_{D}$, where the three contributions are associated with the respective terms of Eq.~(\ref{Eq:Vstatuni}), we have
\begin{gather*}
V_{E1}=
\frac{\mathcal{E}}{2}
\left(v_{E1}\e^{-i\omega t}+v_{E1}^\dag\e^{+i\omega t}\right),
\\
V_{M1}=
\frac{\mathcal{B}}{2}
\left(v_{M1}\e^{-i\omega t}+v_{M1}^\dag\e^{+i\omega t}\right),
\\
V_{D}=
\left(\frac{\mathcal{B}}{2}\right)^2
\left(v_{D}+w_{D}\e^{-i2\omega t}+w_{D}^\dag\e^{+i2\omega t}\right),
\end{gather*}
where the dagger indicates Hermitian conjugation. Dependencies on the field amplitudes and time are given explicitly in the above expressions, while the remaining operators are given by
\begin{gather*}
v_{E1}=-\bm{\epsilon}\cdot\mathbf{d},
\\
v_{M1}=-\bm{\beta}\cdot\bm{\mu},
\\
v_{D}=\left(e^2/4m\right)
\left[\left(\bm{\beta}\times\mathbf{r}\right)\cdot\left(\bm{\beta}^*\times\mathbf{r}\right)\right],
\\
w_{D}=\left(e^2/8m\right)
\left[\left(\bm{\beta}\times\mathbf{r}\right)\cdot\left(\bm{\beta}\times\mathbf{r}\right)\right].
\end{gather*}
Note that since $\mathbf{d}$, $\bm{\mu}$, and $\mathbf{r}$ are themselves Hermitian, Hermitian conjugation of the above operators merely effects the change $\bm{\epsilon}\rightarrow\bm{\epsilon}^*$ and $\bm{\beta}\rightarrow\bm{\beta}^*$.

The perturbation $V$ is periodic in time. Consequently, we may apply conventional Floquet perturbation theory~\cite{Sam73} to obtain expressions for shifts to the atomic energy levels. There are no shifts linear in the field amplitudes, prompting us to consider shifts quadratic in the field amplitudes. Note that $V_{E1}$ and $V_{M1}$ themselves are linear in the field amplitudes, requiring perturbation theory to second order to capture shifts quadratic in the field amplitudes. $V_{D}$, on the other hand, is already quadratic in $\mathcal{B}$ and thus only requires evaluation at first order. The level shifts quadratic in the field amplitudes are
\begin{widetext}
\begin{gather*}
\delta E^{(E1+E1)}_a=
\left(\frac{\mathcal{E}}{2}\right)^2
\sum_b\left(
\frac{
\langle a|v_{E1}|b\rangle
\langle b|v_{E1}^\dag|a\rangle
}{E_a-E_b-\hbar\omega}
+\frac{
\langle a|v_{E1}^\dag|b\rangle
\langle b|v_{E1}|a\rangle
}{E_a-E_b+\hbar\omega}
\right),
\\
\delta E^{(M1+M1)}_a=
\left(\frac{\mathcal{B}}{2}\right)^2
\sum_b\left(
\frac{
\langle a|v_{M1}|b\rangle
\langle b|v_{M1}^\dag|a\rangle
}{E_a-E_b-\hbar\omega}
+\frac{
\langle a|v_{M1}^\dag|b\rangle
\langle b|v_{M1}|a\rangle
}{E_a-E_b+\hbar\omega}
\right),
\\
\delta E^{(E1+M1)}_a=
\frac{\mathcal{E}\mathcal{B}}{4}
\sum_b\left(
\frac{
\langle a|v_{E1}|b\rangle
\langle b|v_{M1}^\dag|a\rangle
}{E_a-E_b-\hbar\omega}
+\frac{
\langle a|v_{M1}|b\rangle
\langle b|v_{E1}^\dag|a\rangle
}{E_a-E_b-\hbar\omega}
+\frac{
\langle a|v_{E1}^\dag|b\rangle
\langle b|v_{M1}|a\rangle
}{E_a-E_b+\hbar\omega}
+\frac{
\langle a|v_{M1}^\dag|b\rangle
\langle b|v_{E1}|a\rangle
}{E_a-E_b+\hbar\omega}
\right),
\\
\delta E_a^{(D)}=
\left(\frac{\mathcal{B}}{2}\right)^2
\langle a|v_{D}|a\rangle,
\end{gather*}
\end{widetext}
where $|a\rangle$ and $E_a$ represent atomic states and energies. Superscripts identify the different contributions to the level shift $\delta E_a$.

The second-order shift $\delta E_a^{(E1+M1)}$ arises from one $V_{E1}$ interaction and one $V_{M1}$ interaction. Since $V_{E1}$ is an odd-parity operator and $V_{M1}$ is an even-parity operator, this contribution vanishes if the atomic states are states of definite parity. For Rydberg states, however, opposite parity states may be degenerate, or at least practically so. Consequently, we do not assume that the atomic states to be employed in the above expressions are states of definite parity.

It is well-known that the Stark shift, $\delta E_a^{(E1+E1)}$, can be partitioned into scalar, vector, and (rank-2) tensor contributions. Such partitioning can be performed for the other shifts as well. We consider the general vector recoupling formula 
\begin{align}
\left\{\mathbf{a}\otimes\mathbf{u}\right\}_k
\cdot\left\{\mathbf{b}\otimes\mathbf{v}\right\}_k
={}&
\left(2k+1\right)
\sum_K
\left\{
\begin{array}{ccc}
1 & 1 & k \\
1 & 1 & K
\end{array}
\right\}
\nonumber\\
&\times
\left(\left\{\mathbf{a}\otimes\mathbf{b}\right\}_K
\cdot\left\{\mathbf{u}\otimes\mathbf{v}\right\}_K\right),
\label{Eq:recouple}
\end{align}
valid if all components of $\mathbf{u}$ and $\mathbf{b}$ commute. Here the factor immediately following the summation symbol is a Wigner $6j$ symbol, $\left\{\mathbf{a}\otimes\mathbf{b}\right\}_k$ denotes the rank-$k$ irreducible tensor product of the vectors $\mathbf{a}$ and $\mathbf{b}$, and the dot signifies the dot product between irreducible tensors of the same rank~\cite{VarMosKhe88}. For $k=0,1$, the tensor product $\left\{\mathbf{a}\otimes\mathbf{b}\right\}_k$ can be related to the familiar dot and cross products involving vectors, $\left\{\mathbf{a}\otimes\mathbf{b}\right\}_0=-\left(1/\sqrt{3}\right)\left(\mathbf{a}\cdot\mathbf{b}\right)$ and $\left\{\mathbf{a}\otimes\mathbf{b}\right\}_1=\left(i/\sqrt{2}\right)\left(\mathbf{a}\times\mathbf{b}\right)$. In Eq.~(\ref{Eq:recouple}), $K$ runs over the values $K=0,1,2$. For our purposes, the vectors $\mathbf{a}$ and $\mathbf{b}$ appearing in Eq.~(\ref{Eq:recouple}) identify with the polarization vectors $\bm{\epsilon}$, $\bm{\beta}$, or their complex conjugates, with $\mathbf{u}$ and $\mathbf{v}$ being independent of the polarization. In the present work, we are interested in shifts due to BBR, which requires averaging the plane-wave shifts over all possible polarizations. To this end, recoupling according to Eq.~(\ref{Eq:recouple}) is advantageous. Namely, since the perturbing BBR environment is isotropic, the nonscalar $(K\neq0)$ contributions necessarily average to zero. With this in mind, we rewrite Eq.~(\ref{Eq:recouple}) in a form appropriate for our purposes,
\begin{gather*}
\left[\left(\mathbf{a}\cdot\mathbf{u}\right)
\left(\mathbf{b}\cdot\mathbf{v}\right)\right]_\mathrm{scalar}
=(1/3)\left(\mathbf{a}\cdot\mathbf{b}\right)
\left(\mathbf{u}\cdot\mathbf{v}\right),
\\
\left[\left(\mathbf{a}\times\mathbf{u}\right)\cdot
\left(\mathbf{b}\times\mathbf{v}\right)\right]_\mathrm{scalar}
=(2/3)\left(\mathbf{a}\cdot\mathbf{b}\right)
\left(\mathbf{u}\cdot\mathbf{v}\right),
\end{gather*}
where the top and bottom expressions are for $k=0$ and $k=1$, respectively. Here $[\cdots]_\mathrm{scalar}$ indicates that recoupling is to be performed in accordance with Eq.~(\ref{Eq:recouple}), with only the scalar ($K=0$) term being retained.

To demonstrate application of the above formulas, we consider the first term of $\delta E_a^{(E1+E1)}$. The numerator of this term is $\langle a|\bm{\epsilon}\cdot\mathbf{d}|b\rangle\langle b|\bm{\epsilon}^*\cdot\mathbf{d}|a\rangle=\left(\bm{\epsilon}\cdot\langle a|\mathbf{d}|b\rangle\right)\left(\bm{\epsilon}^*\cdot\langle b|\mathbf{d}|a\rangle\right)$. Keeping only the scalar term upon recoupling, we find
\begin{gather*}
\left[\langle a|\bm{\epsilon}\cdot\mathbf{d}|b\rangle\langle b|\bm{\epsilon}^*\cdot\mathbf{d}|a\rangle\right]_\mathrm{scalar}
=(1/3)\left|\langle a|\mathbf{d}|b\rangle\right|^2,
\end{gather*}
where we used $\langle b|\mathbf{d}|a\rangle=\langle a|\mathbf{d}|b\rangle^*$ and the fact that $\bm{\epsilon}$ is a unit vector, $\bm{\epsilon}\cdot\bm{\epsilon}^*=1$.

The case of $\delta E_a^{(E1+M1)}$ requires special consideration. Considering the numerator of the first term, we have
\begin{align*}
\left[\langle a|\bm{\epsilon}\cdot\mathbf{d}|b\rangle\langle b|\bm{\beta}^*\cdot\bm{\mu}|a\rangle\right]_\mathrm{scalar}
={}&(1/3)\left(\bm{\epsilon}\cdot\bm{\beta}^*\right)
\\&\times
\left(\langle a|\mathbf{d}|b\rangle\cdot\langle b|\bm{\mu}|a\rangle\right).
\end{align*}
Meanwhile, $\bm{\epsilon}\cdot\bm{\beta}^*=i\mathcal{A}$, where $\mathcal{A}$ is the degree of circular polarization. $\mathcal{A}$ satisfies $-1\leq\mathcal{A}\leq1$, where the extreme values correspond to left- and right-hand-circular polarization, respectively, and $\mathcal{A}=0$ corresponds to linear polarization. More generally, $\left|\mathcal{A}\right|$ specifies the eccentricity, while the sign of $\mathcal{A}$ specifies the handedness. Since BBR is unpolarized with no net-handedness, this term necessarily averages to zero. This is the case for the other terms of $\delta E_a^{(E1+M1)}$ as well, such that $\delta E_a^{(E1+M1)}$ does not contribute to a BBR shift even if the atomic states are not states of definite parity.

Excluding $\delta E_a^{(E1+M1)}$ and retaining only the scalar contributions for the remaining shifts, we find
\begin{gather*}
\delta E^{(E1+E1)}_a=
\left(\frac{\mathcal{E}}{2}\right)^2
\left[\frac{2}{3\hbar}\sum_b
\left|\left\langle a|\mathbf{d}|b\right\rangle\right|^2
\frac{\omega_{ab}}{\omega_{ab}^2-\omega^2}
\right],
\\
\delta E^{(M1+M1)}_a=
\left(\frac{\mathcal{B}}{2}\right)^2
\left[\frac{2}{3\hbar}\sum_b
\left|\left\langle a|\bm{\mu}|b\right\rangle\right|^2
\frac{\omega_{ab}}{\omega_{ab}^2-\omega^2}
\right],
\\
\delta E^{(D)}_a=
\left(\frac{\mathcal{B}}{2}\right)^2
\left[\frac{e^2}{6m}\langle a|r^2|a\rangle\right],
\end{gather*}
where we have introduced $\omega_{ab}=\left(E_a-E_b\right)/\hbar$ for brevity. For $\delta E^{(E1+E1)}_a$, the factor in square brackets is the familiar scalar Stark polarizability, excluding a negative sign. For $\delta E^{(M1+M1)}_a$ and $\delta E^{(D)}_a$, the factors in square brackets together give the analogous scalar Zeeman polarizability, excluding a negative sign.

The contribution $\delta E^{(D)}_a$ is positive definite. For the ground state in the dc limit, the contribution $\delta E^{(M1+M1)}_a$ is negative definite. The former identifies with diamagnetism, while the latter identifies with paramagnetism. More generally, the sign of $\delta E^{(M1+M1)}_a$ is not definite. Nevertheless, we refer to $\delta E^{(D)}_a$ exclusively as the diamagnetic contribution to the second order Zeeman shift, with this term being independent of the frequency.

Expressions thus far assume monochromatic radiation. However, BBR has a spectral distribution given by Planck's law,
\begin{gather*}
u(\omega,T)=\frac{\hbar\omega^3}{\pi^2c^3}\frac{1}{e^{\hbar\omega/k_BT}-1}.
\end{gather*}
Here $u(\omega,T)$ is the spectral energy density, which depends on frequency $\omega$ and temperature $T$. To obtain the BBR shift, we make the following associations for the field amplitudes
\begin{gather*}
\left(\frac{\mathcal{E}}{2}\right)^2\rightarrow\frac{u(\omega,T)}{2\varepsilon_0}d\omega,
\\
\left(\frac{\mathcal{B}}{2}\right)^2\rightarrow\frac{u(\omega,T)}{2\varepsilon_0c^2}d\omega,
\end{gather*}
and integrate over all frequencies ($d\omega$ being the differential frequency element). Poles in the integration are appropriately handled by taking the Cauchy principal value~\cite{FarWin81}.

For a high-lying Rydberg state $a$, the electric dipole matrix element $\left\langle a|\mathbf{d}|b\right\rangle$ is only appreciable for states $b$ that are energetically close to the state $a$. Supposing only states satisfying $\left|\omega_{ab}\right|\ll k_BT/\hbar$ contribute appreciably to the BBR Stark shift, the following approximation can be made within the integrand,
\begin{gather}
\frac{u(\omega,T)}{\omega_{ab}^2-\omega^2}\approx-\frac{u(\omega,T)}{\omega^2}.
\label{Eq:highlyingBBRsub}
\end{gather}
In this case, the integral can be performed analytically and the Thomas-Reiche-Kuhn sum rule can be invoked for the sum over states $b$. This yields the result
%\begin{gather*}
%\delta E^{(E1+E1)}_a\approx
%\frac{\pi\alpha\left(k_BT\right)^2}{3mc^2}
%\sum_bf_{ba},
%\end{gather*}
%where the $f_{ba}$ are conventional oscillator strengths,
%\begin{gather*}
%f_{ba}=\frac{2m}{3\hbar}\left|\left\langle b|\mathbf{r}|a\right\rangle\right|^2\omega_{ba}.
%\end{gather*}
%Meanwhile, for a single-electron system, the Thomas-Reiche-Kuhn sum rule~\cite{} insists that
%$\sum_bf_{ba}=1$, with a corresponding BBR Stark shift of
\begin{gather}
\delta E^{(E1+E1)}_a\approx
\frac{\pi\alpha\left(k_BT\right)^2}{3mc^2},
\label{Eq:FWasy}
\end{gather}
referred to as the ``high-$n$ approximation'' by Farley and Wing~\cite{FarWin81}. Here $n$ refers to the principal quantum number of the Rydberg electron. For $T=300$~K, Eq.~(\ref{Eq:FWasy}) evaluates to $\delta E^{(E1+E1)}_a/h\approx 2.42$~kHz. In their work, Farley and Wing considered different Rydberg series of the alkali metal atoms and helium, evaluating the BBR Stark shift for values of $n$ up to 30. The progression of the BBR Stark shift with respect to $n$ is depicted in several insightful figures. The progression is generally nonmonotonic and has varying behavior among the different species and series, but in all cases a tendency towards the high-$n$ approximation is evident for the largest values of $n$.

%The BBR Stark shift is expected to dominate over the BBR Zeeman shift. However, for transitions between atomic states, only the differential shift between energy levels leads to a shift in the transition frequency. As the BBR Stark shift to the energy levels is given, at least approximately, by a simple universal expression for high-lying Rydberg states, Eq.~(\ref{Eq:FWasy}), a high degree of cancellation may be expected between these states. As a consequence, BBR shifts to the transition frequencies may not be dominated by the BBR Stark shift.

Generally speaking, $\delta E^{(M1+M1)}_a$ is expected to be negligible. Eigenvalues of $H_0$ are specified by the pair of quantum numbers $n$ and $l$, where $l$ is the conventional quantum number specifying orbital angular momentum. Meanwhile, the magnetic dipole operator $\bm{\mu}$ is diagonal in both $n$ and $l$. It follows that $\left|\left\langle a|\bm{\mu}|b\right\rangle\right|^2\omega_{ab}=0$ and, consequently, $\delta E^{(M1+M1)}_a=0$. However, effects not explicitly incorporated in $H_0$, such as the spin-orbit interaction, can lift the degeneracy among states of a given $n$ and $l$, enabling a nonzero shift. To give a quantitative example, we consider the $p$ ($l=1$) Rydberg series of Rb, with fine structure components $p_{1/2}$ and $p_{3/2}$ resulting from the spin-orbit interaction. Once again, approximation~(\ref{Eq:highlyingBBRsub}) may be used in the integrand, with the integral over $\omega$ being performed analytically. For the remaining sum over states, we may take
\begin{gather*}
\left|\sum_b\left|\left\langle a|\bm{\mu}|b\right\rangle\right|^2
\frac{\omega_{ab}}{2\pi}\right|
\approx\left(\frac{4}{3}\frac{\mu_B^2}{2j+1}\right)\left[\frac{8.6\times10^{13}~\text{Hz}}{(n-2.7)^3}\right],
\end{gather*}
where $j=1/2,3/2$ specifies the fine structure component under consideration and the factor in square brackets approximates the splitting between the fine structure components. The numerical value of 2.7 is the approximate quantum defect for the $p$ series. Evaluating $|\delta E^{(M1+M1)}_a/h|$ yields $\lesssim15~\mu\text{Hz}$ for all $n$ (i.e., $n\geq 5$) and $\lesssim 36~\text{nHz}$ for $n\geq20$. This is consistent with our general expectation that $\delta E^{(M1+M1)}_a$ may be neglected, in spite of effects that may lift the degeneracy among states of a given $n$ and $l$.

Finally, we consider the diamagnetic term $\delta E^{(D)}_a$. In this case, the integral over $\omega$ may be performed analytically, yielding
\begin{gather*}
\delta E_a^{(D)}=\frac{\pi^3\alpha\left(k_BT\right)^4}{45m\hbar^2c^4}\langle a|r^2|a\rangle.
\end{gather*}
As for the expectation value of $r^2$, we introduce the states $|nlm_l\rangle$, where $m_l$ is the conventional quantum number specifying the projection of orbital angular momentum onto the quantization axis. Since $r^2$ only acts on the spatial degrees of freedom, we omit the analogous quantum number $m_s$ for the spin angular momentum. In the case of hydrogen, the expectation value of $r^2$ is given by a simple analytical expression~\cite{GalCoo79}
\begin{gather}
\langle nlm_l|r^2|nlm_l\rangle=\frac{a_B^2}{2}n^2\left[5n^2-3l(l+1)+1\right].
\label{Eq:r2hydrogen}
\end{gather}
Being an expectation value that depends predominantly on the electron probability density away from the core region, we anticipate that this formula holds, to good approximation, for more general Rydberg atoms so long as the principal quantum number on the right-hand-side is replaced with the effective principal quantum number (principal quantum number minus quantum defect). Assuming $n\lesssim 100$, the above formulas together imply
\begin{gather*}
\delta E^{(D)}_a/h\lesssim 20~\text{Hz},
\end{gather*}
indicating that the BBR Zeeman shift is two or more orders of magnitude below the BBR Stark shift. In contrast to the BBR Stark shift, however, the BBR Zeeman shift scales steeply with $n$ rather than asymptotically approaching a constant value.

\section{A quantitative example}

Here we consider a specific transition between Rydberg states of Rb, with the pair of states being specified below. Our reason for choosing this particular transition is two-fold. Firstly, there is a practical interest in this transition. As noted in the Introduction, Ramos~{\it et al.}~\cite{RamMooRai17} have proposed a Rydberg-atom experiment that aims to determine the Rydberg constant. The proposal focuses on this transition, with the unperturbed transition frequency being nearly proportional to the Rydberg constant. Secondly, Ramos~{\it et al.}\ have already performed calculations of the BBR Stark shift for the two states involved in the transition, alleviating the need for such calculations here. The results presented here merely serve as a quantitative example, with other transitions potentially being of interest for high-precision spectroscopy experiments with Rydberg atoms.

For states with large values of $l$, the inner centrifugal barrier strongly inhibits the electron from penetrating the core region. As a consequence, these states are well described by hydrogenic results, with the states being effectively degenerate for a given value of $n$. External dc fields can be used to lift this degeneracy and define the ``good'' states of the manifold. The pair of states considered by Ramos~{\it et al.}\ identify with states of the ``parabolic'' basis $|nn_1n_2m_l\rangle$. This basis is related to the ``spherical'' basis $|nlm_l\rangle$ according to~\cite{RamMooRai17,Gal94}
\begin{gather*}
|nn_1n_2m_l\rangle=\sum_lC_{lm_l}^{n_1n_2}|nlm_l\rangle,
\end{gather*}
where the coefficients $C_{lm_l}^{n_1n_2}$ are
\begin{align*}
C_{lm_l}^{n_1n_2}=
{}&(-1)^{(1-n+m_l+n_1-n_2)/2+l}\sqrt{2l+1}.
\\&\times
\left(\begin{array}{ccc}
\frac{n-1}{2} & \frac{n-1}{2} & l \\
\frac{m_l+n_1-n_2}{2} & \frac{m_l-n_1+n_2}{2} & -m_l
\end{array}\right).
\end{align*}
The factor on the second line here is a Wigner 3$j$ symbol. In the parabolic basis, the principal quantum number is redundant, satisfying $n=n_1+n_2+|m_l|+1$. Specifically, the two states of interest are
\begin{gather*}
|51,0,0,50\rangle=|51,50,50\rangle,
\\
|53,1,1,50\rangle=\sqrt{\frac{51}{103}}|53,50,50\rangle-\sqrt{\frac{52}{103}}|53,52,50\rangle.
\end{gather*}
In each case here, the state is expressed in terms of the parabolic basis on the left-hand-side and the spherical basis on the right-hand-side. The first state, with $|m_l|=n-1$ and for which there is a one-to-one relation between basis states, is a so-called circular state. The other state may be regarded as a near-circular state.

To evaluate the BBR Zeeman shift, we require the expectation value of $r^2$ in the parabolic basis. The expectation value of $r^2$ in the spherical basis is given by Eq.~(\ref{Eq:r2hydrogen}). For these purposes, the quantum defect is negligible, permitting direct use of the hydrogenic result. Since $r^2$ is a scalar operator, it does not mix states of different $l$ and $m_l$. Consequently, the expectation value of $r^2$ in the parabolic basis satisfies
\begin{gather*}
\langle nn_1n_2m_l|r^2|nn_1n_2m_l\rangle
=
\sum_l\left|C_{lm_l}^{n_1n_2}\right|^2
\langle nlm_l|r^2|nlm_l\rangle.
\end{gather*}
With this, we have the expressions necessary to evaluate the BBR Zeeman shift.

Table~\ref{Tab:results} presents results for the BBR Stark shift and the BBR Zeeman shift for the pair of Rydberg states considered by Ramos~{\it et al.} Results for the BBR Stark shift are taken from Refs.~\cite{RamMooRai17,Ramosthesis}, where values close to the asymptotic value given by Eq.~(\ref{Eq:FWasy}) are observed. As expected, we see that the BBR Stark shift dominates over the BBR Zeeman shift. However, for shifts to the transition frequency, we find that the BBR Zeeman shift is about six times larger than the BBR Stark shift, with opposite sign. Combined, these correspond to a $1.1\times10^{-12}$ fractional shift to the transition frequency. For comparison, the latest \codata{} recommended value for the Rydberg constant has a relative uncertainty of $1.1\times10^{-12}$~\cite{CODATA2022online}.

\begin{table}[t]
\caption{BBR shifts for Rydberg states of Rb at 300~K, partitioned into Stark and Zeeman contributions. The BBR Stark shift data is from Refs.~\cite{RamMooRai17,Ramosthesis}. The bottom line (``transition'') gives the differential shift between the two levels. For reference, Eq.~(\ref{Eq:FWasy}) evaluates to $2416.666$~Hz. All quantities are in units of hertz (i.e., for the states, the entries correspond to $\delta E_a/h$ in units of hertz).}
\label{Tab:results}
\begin{ruledtabular}
\begin{tabular}{lcc}
$|n,n_1,n_2,m_l\rangle$	& Stark			& Zeeman	\\
\hline
$|51,0,0,50\rangle$		& $2416.661$	& $0.532$	\\
$|53,1,1,50\rangle$		& $2416.640$	& $0.653$	\\
\hline
transition				& $-0.021$		& $0.121$
\end{tabular}
\end{ruledtabular}
\end{table}

Aside from level shifts, we note that BBR also induces transitions between Rydberg states, shortening the effective lifetime of the states. To mitigate this effect, Ramos~{\it et al.}\ suggested performing the spectroscopy in a cryogenic environment. An additional benefit offered by the cryogenic environment is suppressed BBR shifts, including both Stark and Zeeman contributions.

\section{Core contributions and higher-order Stark effects}

As noted at the start of Sec.~\ref{Sec:derive}, we use a single-active electron model. This neglects direct contributions of the core to the BBR Stark and BBR Zeeman shifts. In principle, these core contributions should be added to the level shifts. However, as they amount to a common energy offset for the Rydberg states, they are not of consequence for transitions between the Rydberg states. Nevertheless, we consider them here for the sake of completeness. The dc Stark polarizability of the Rb ionic core was measured in Ref.~\cite{BerSacGal20}. Using that result, together with the fact that the core excitations have transition frequencies far exceeding $k_BT/\hbar$, we calculate a core contribution to the BBR Stark shift of $\delta E^{(E1+E1)}_\mathrm{core}/h=-78.5~\text{mHz}$. Notably, this is within the significant digits provided in Table~\ref{Tab:results}. Meanwhile, the BBR Zeeman shift is proportional to the expectation value of $r^2$. In the multi-electron case, this must be extended to include a summation over all electrons, $r^2\rightarrow\sum_ir_i^2$. We estimate that the core contribution to the expectation value is $\lesssim\!N_ca_B^2$, where $N_c$ is the number of core electrons ($N_c=36$ for Rb). For the states in Table~\ref{Tab:results}, the fractional correction to the BBR Zeeman shift is thus $\lesssim\!5\times10^{-6}$, being justifiably neglected.

The focus of this work is the BBR Zeeman shift. Stark effects are included for context, as well as to consider a possible mixed effect (i.e., $\delta E^{(E1+M1)}_a$). Higher order Stark effects (e.g., fourth order in $V_{E1}$) are not considered here, though they could potentially be relevant for high-precision experiments with Rydberg atoms as well.

\section{Conclusion}

We have considered the BBR Zeeman shift in Rydberg atoms. The BBR Zeeman shift is attributed almost entirely to the diamagnetic contribution, which scales steeply with the principal quantum number of the Rydberg electron ($\sim\!n^4$). Nevertheless, for practical values of $n$, the BBR Stark shift remains the dominant contribution to the level shifts. However, in contrast to the BBR Zeeman shift's steep scaling with respect to $n$, the BBR Stark shift asymptotically approaches a constant value given by a universal expression. We find that for transitions between Rydberg states, where only the differential shift between levels is of concern, the BBR Zeeman shift can surpass the BBR Stark shift. This could be relevant for future high-precision experiments that leverage the special features of Rydberg atoms.

~

\begin{acknowledgments}
We thank Alejandra Collopy and Travis Briles for their careful reading of the manuscript. This work was supported by the National Institute of Standards and Technology Innovations in Measurement Sciences program.
\end{acknowledgments}

%\bibliography{Rydberg_diamagnetic_BBR_biblio}

%apsrev4-2.bst 2019-01-14 (MD) hand-edited version of apsrev4-1.bst
%Control: key (0)
%Control: author (8) initials jnrlst
%Control: editor formatted (1) identically to author
%Control: production of article title (0) allowed
%Control: page (0) single
%Control: year (1) truncated
%Control: production of eprint (0) enabled
%

\end{document}